\documentclass[a4paper,12pt]{article}
\usepackage{amsmath}
\usepackage{citehack}
\usepackage{bm}
\usepackage{graphicx}
\usepackage{epstopdf}
\usepackage{color}
\usepackage{comment}
\usepackage{hyperref}
\usepackage{float}
\usepackage{tikz, xcolor, pgfplots}
\usetikzlibrary{shapes, arrows}
\usepackage{multirow}

\begin{document}

\title{Molecular dynamic simulation of water vapor interaction 
with various types of pores using hybrid computing structures
\thanks{topic JINR LIT No. 05-6-1118-2014/2019, protocol No. 4596-6-17/19. HybriLIT resources.}}
\author{V.V. Korenkov$^1$,  E.G. Nikonov$^1$,  M. Popovi\v{c}ov\'a$^2$\\
\begin{minipage}{10cm}
\begin{center}
\small\em
\vspace{4mm}
$^1$Joint Institute for Nuclear Research,\\ 141980 Dubna, Moscow Region, Russia\\ email: korenkov@jinr.ru, e.nikonov@jinr.ru \\
\vspace{4mm}
$^2$University of Pre\v{s}ov,\\ str. Kon\v{s}tantinova 16, 080 01 Pre\v{s}ov,  Slovakia\\ email: maria.popovicova@unipo.sk
\end{center}
\end{minipage}
} 
\date{}
\maketitle 

\begin{abstract}
Theoretical and experimental investigations of water vapor interaction with porous materials are very needful for various fields of science and technology. Not only studies of the interaction of water vapor and porous material as a continuous medium, but also the study of the interaction of water vapor with individual pore is very important in these researches. Mathematical modelling occupies an important place in these investigations. Conventional approaches to solve problems of mathematical research of the processes of interaction of water vapor with individual pore are the following. The first approach is based on the use of diffusion equation for description of interaction of water vapor with a pore. It is so called macro approach. The second approach is based on various particle methods like, for example, molecular dynamics (MD). These methods essentially consider the micro-structure of the investigated system consisting of water vapor and a pore. This second approach can be called a micro approach.

At the macro level, the influence of the arrangement structure of individual pores on the processes of water vapor interaction with porous material as a continuous medium is studied. At the micro level, it is very interesting to investigate the dependence of the characteristics of the water vapor interaction with porous media on the geometry and dimensions of the individual pore. Both approaches require the most efficient calculation methods as far as possible with the current level of development of computational technologies. Usage of efficient calculation methods is necessary because the degree of approximation for simulating system is largely determined by the dimensionality of the system of equations being solved at every time step. Number of time steps is also quite large. 

In this work, a study of efficiency of various implementations algorithms for MD simulation of water vapor interaction with individual pore is carried out. A great disadvantage of MD is its requirement of a relatively large computational effort and long time in simulations. These problems can be drastically reduced by parallel calculations. In this work we investigate dependence of time required for simulations on different parameters, like number of particles in the system, shape of pores, and so on. The results of parallel calculations are compared with the results obtained by serial calculations.
\end{abstract}

\section{Introduction}

One of the most important problem in numerical simulation based on molecular dynamics or Monte-Carlo approach of many particle systems  is the need to use huge computing resources to obtain more or less realistic simulation results. A system of water vapor and a pore is an example of such many particle systems. Theoretical and experimental investigations of water vapor interaction with porous materials are very needful for various fields of science and technology. Not only studies of the interaction of water vapor and porous material as a continuous medium, but also the study of the interaction of water vapor with individual pore is very important in these researches. Mathematical modelling occupies an important place in these investigations. Conventional approaches to solve problems of mathematical research of the processes of interaction of water vapor with individual pore are the following. The first approach is based on the use of diffusion equation for description of interaction of water vapor with a pore. It is so called macro approach. The second approach is based on various particle methods like, for example, molecular dynamics (MD). These methods essentially consider the micro-structure of the investigated system consisting of water vapor and a pore. This second approach can be called a micro approach.

At the macro level, the influence of the arrangement structure of individual pores on the processes of water vapor interaction with porous material as a continuous medium is studied. At the micro level, it is very interesting to investigate the dependence of the characteristics of the water vapor interaction with porous media on the geometry and dimensions of the individual pore. Both approaches require the most efficient calculation methods as far as possible with the current level of development of computational technologies. Usage of efficient calculation methods is necessary because the degree of approximation for simulating system is largely determined by the dimensionality of the system of equations being solved at every time step. Number of time steps is also quite large. 

In this work, a study of efficiency of various implementations algorithms for MD simulation of water vapor interaction with individual pore is carried out. A great disadvantage of MD is its requirement of a relatively large computational effort and long time in simulations. These problems can be drastically reduced by parallel calculations. In this work we investigate dependence of time required for simulations on different parameters, like number of particles in the system, shape of pores, and so on. The results of parallel calculations are compared with the results obtained by serial calculations. Two-dimensional and three-dimensional models of the pore are used for comparative analysis of parallel and serial calculations.

\section{Molecular dynamics model}
\label{mdm}
In classical molecular dynamics, the behavior of an individual particle is described by the Newton equations of motion \cite{Gould}, which can be written in the following form
\begin{equation}\label{a} 
  m_i\frac{d^2 \vec{r_i}}{dt^2}=\vec{f_i},
\end{equation}
\noindent
where $i \ - $ a particle number, $(1\leq i \leq N)$, $N \ - $ the total number of particles, $m_i \ - $ particle mass, $\vec{r_i}\ - $ coordinates of position,  $\vec{f_i} \ - $ the resultant of all forces acting on the particle. This resultant force has the following representation
\begin{equation}\label{b} 
  \vec{f_i} = -\frac{\partial U(\vec{r_1},\ldots,\vec{r_N})}{\partial \vec{r_i}} + \vec{f_i}^{ex},
\end{equation}
where $U \ - $ the potential of particle interaction, $\vec{f_i}^{ex} - $ a force caused by external fields. 
For a simulation of particle interaction, we use the Lennard-Jones potential \cite{LJ} with $\sigma = 3.17 \mbox{\AA}$ and $\varepsilon = 6.74\cdot 10^{-3}$ eV. It is the most used to describe the evolution of water in liquid and saturated vapor form. Equations of motion 
(\ref{a}) 
were integrated by Velocity Verlet method \cite{Verlet1967}.
Berendsen thermostat \cite{Berendsen1984} is used for temperature calibration and control. The coefficient of the velocity recalculation $\lambda(t)$ at every time step $t$ depends on the so called ''rise time'' of the thermostat $\tau_B$ which belongs to the interval $ [0.1,2] \ \mbox{psec}$. 
$\tau_B$ describes strength of the coupling of the system to a hypothetical heat bath. For increasing $\tau_B$, the coupling weakens, i.e. it takes longer to achieve given temperature $T_0$ from current temperature $T(t).$
The Berendsen algorithm is simple to implement and it is very efficient for reaching the desired temperature from far-from-equilibrium configurations.

Initial concentrations were obtained from the density of water vapor at the appropriate pressure and density at a given temperature using known tabulated data. The pressure in the pore was controlled using the formula based on virial equation \cite{Frenkel2002}.
$$
P = \frac{1}{3V}\left(\left\langle 2K\right\rangle-\left\langle\sum\limits_{i<j} r_{ij}\cdot f\left(r_{ij}\right)\right\rangle\right).
$$
Here $V$ is the pore volume, $\left\langle 2K\right\rangle$ is the doubled kinetic energy averaged over the ensemble, $f\left(r_{ij}\right)$ is the force between  particles $i$ and $j$ at a distance $r_{ij}$.

\section{Computational algorithm for molecular dynamic simulation}
For molecular dynamic simulation we used the code written in \textbf{CUDA C}. The program does not require a lot of memory. We only keep co-ordinates, speeds and forces for each particle. One of the main problems of molecular dynamic simulation is a large number of particles and time steps. Therefore it is necessary to use parallel calculations. The code for our simulations was implemented  on  heterogeneous  computing  cluster  \textit{HybriLIT}.

The code contains four functions that are paralleled and which are performed on the \textbf{GPU}. This is a function for calculating the forces (i.e., acceleration) for individual particles, which calculates the interactions between all particles (F1). There are two functions to calculate new coordinates and speeds for each particle. We need two functions to calculate them because we need forces acting on particles at two different time moments(F2 and F3).  Finally, we use the Berendsen thermostat in the program that runs parallel to the \textbf{GPU} too (F4).

We use natural parallelism for molecular dynamic simulations. The force calculations and velocity/position updates can be done simultaneously for all particles. 

There are two basic ideas how to achieve parallelism. The goal in each is to divide computations evenly across the processors so as to extract maximum effect.

In the first class of methods a subgroup of particles is assigned to each processor. This method is called an particle-decomposition of the workload. The processor performs all calculations on its particles no matter where they move in the simulation domain.

The second group of methods is called a spatial decomposition of the workload. It means that parts of the physical simulation domain is assigned to each processor. Each processor only works with the particles in its subdomain.

Our program uses an particle-decomposition method. One command provides processing of a large amount of data that depends on how the block  is defined in the program. The pseudo code for all four parallel functions is in Fig. \ref{f1} - \ref{f4}

\begin{figure}[H]
\begin{center}
\begin{verbatim}
__global__void F1( ) { 

int tid = threadIdx.x + blockIdx.x*blockDim.x; 
while(tid < N) { 
\end{verbatim}

\begin{flushleft}
// \textbf{Derive
$a_{tid}(t + \Delta t)$ 
from the interaction potential using $r(t + \Delta t)$}
\end{flushleft}
\begin{verbatim}
tid += blockDim.x*gridDim.x; 
} 
}
\end{verbatim}
\end{center}
\vspace{-0.8cm}
\caption{The function for the calculation of acceleration, respected forces of each particle on the device}
\label{f1}
\end{figure}
\begin{figure}[H]
\begin{center}
\begin{verbatim}
__global__void F2( ) { 

int tid = threadIdx.x + blockIdx.x*blockDim.x; 
while(tid < N) { 
\end{verbatim}

\begin{flushleft}
/*
 \textbf{Calculate
$r_{tid}(t + \Delta t) = r_{tid}(t) + v_{tid}(t)*\Delta t + 0.5*a_{tid}(t)*\Delta t^2$
$v_{tid}(t + 0.5*\Delta t) = v_{tid}(t) + 0.5*a_{tid}(t)*\Delta t$
}
*/
\end{flushleft}
\begin{verbatim}
tid += blockDim.x*gridDim.x; 
} 
}
\end{verbatim}
\end{center}
\vspace{-0.8cm}
\caption{The function for calculating a position and first part of velocity for each particle performed on the device}
\label{f2}
\end{figure}
\begin{figure}[H]
\begin{center}
\begin{verbatim}
__global__void F3( ) { 

int tid = threadIdx.x + blockIdx.x*blockDim.x; 
while(tid < N) { 
\end{verbatim}

\begin{flushleft}
//
 \textbf{Calculate
$v_{tid}(t + \Delta t) = v_{tid}(t + 0.5*\Delta t) + 0.5*a_{tid}(t + \Delta t)*\Delta t$}

\end{flushleft}
\begin{verbatim}
tid += blockDim.x*gridDim.x; 
} 
}
\end{verbatim}
\end{center}
\vspace{-0.8cm}
\caption{The function for calculating the second speed fraction for each particle performed on the device}
\label{f3}
\end{figure}
\begin{figure}[H]
\begin{center}
\begin{verbatim}
__global__void F4( ) { 

int tid = threadIdx.x + blockIdx.x*blockDim.x; 
while(tid < N) { 
\end{verbatim}

\begin{flushleft}
//
 \textbf{Calculate
$v_{tid}(t) = v_{tid}(t)*\lambda (t)$}

\end{flushleft}
\begin{verbatim}
tid += blockDim.x*gridDim.x; 
} 
}
\end{verbatim}
\end{center}
\vspace{-0.8cm}
\caption{The function for speed adjustment to ensure the required temperature (The Berendsen thermostat)}
\label{f4}
\end{figure}
\begin{figure}[H]
\begin{center}
\includegraphics[width=1.05\linewidth]{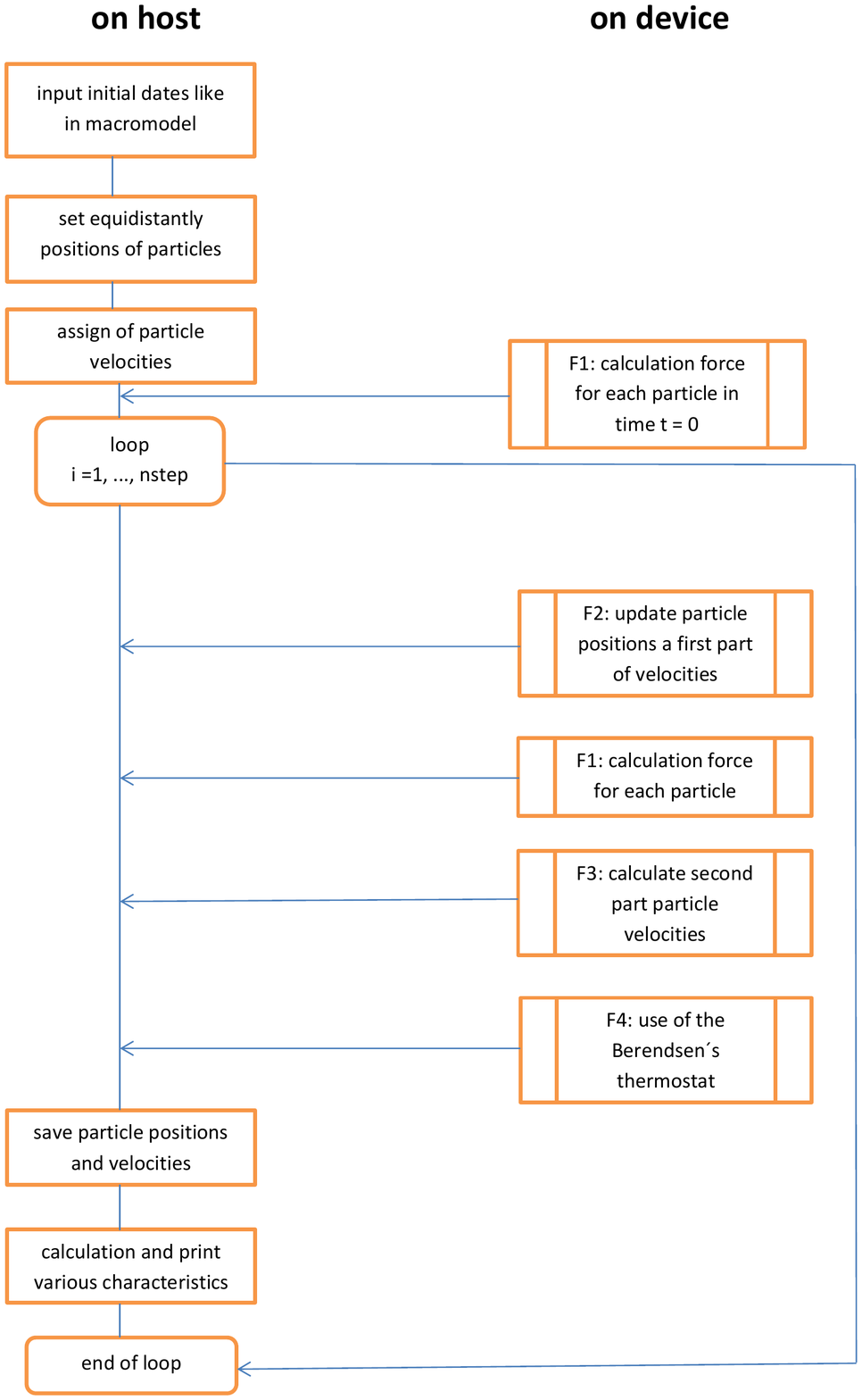}
\end{center}
\vspace{-1cm}
\caption{Computational scheme for molecular dynamic simulation}
\label{VD_hyb}
\end{figure}

Other calculations are performed on the host. General scheme of the calculation algorithm for two- and three-dimensional molecular dynamic simulation is shown in Fig. \ref{VD_hyb}.

In this paper we compare the temporal realization of these four functions $F_1, F_2, F_3, F_4$ on the \textbf{GPU} and the \textbf{CPU}. The total time of parallel computing consists of two parts, that is the time needed directly to calculate on the \textbf{GPU} (pure \textbf{GPU} time) and the time needed to complete these calculations on the \textbf{CPU} because some algorithms performed on the \textbf{GPU} must be completed by the \textbf{CPU}. In this work, total \textbf{GPU} time will indicate the sum of these two times.

\section{2D molecular dynamic simulation}
\begin{figure}[H]
\begin{center}
\begin{tikzpicture}[scale=1.5]
    \draw[ultra thick] (3,0)--(0,0)--(0,1.5)--(3,1.5);
    \filldraw[black] (1.1,0.75) node[anchor=west] {pore};
    \draw[dashed] (3,1.5)--(3,0);
    \draw[dashed][ultra thick] (3,1.5)--(3,3)--(6,3)--(6,-1.5)--(3,-1.5)--(3,0); 
    \filldraw[black] (3.7,0.75) node[anchor=west] {outer space};
    \draw (-0.6,0)--(0,0); 
    \draw (-0.6,1.5)--(0,1.5); 
    \draw (-0.5,-0.1)--(-0.5,1.6); 
    \filldraw[black] (-0.5,0.7) node[anchor=east] {$l_y$};  
    \draw (0,0)--(0,-2.1); \draw (3,-1.5)--(3,-2.1); \draw (6,-1.5)--(6,-2.1); \draw (-0.1,-2)--(6.1,-2);
    \filldraw[black] (1.5,-1.6) node[anchor=north] {$l_x$};
    \filldraw[black] (4.5,-1.6) node[anchor=north] {$k\cdot l_x$};
    \draw (6,-1.5)--(6.7,-1.5); \draw (6,3)--(6.7,3); \draw (6.6,-1.6)--(6.6,3.1);  
    \filldraw[black] (6.7,0.7) node[anchor=west] {$(2\cdot k+1)\cdot l_y$};  
\end{tikzpicture}
\end{center}
\caption{2D pore and outer space.} 
\label{fig1}
\end{figure}
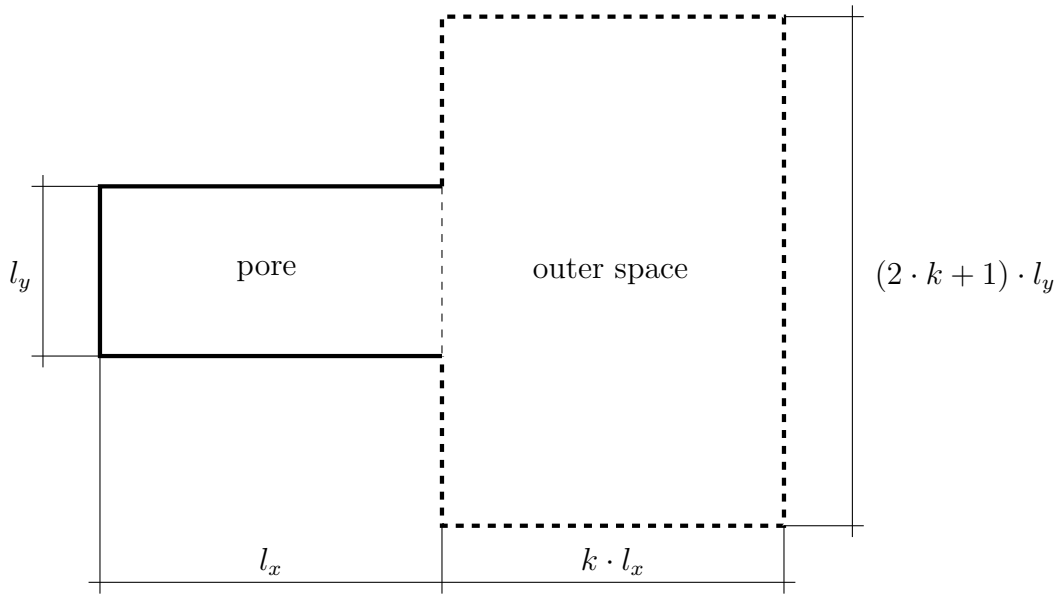

We consider the pore with dimensions  $l_x = 1\mu m$, $l_y = 1\mu m$. The outer space in this micro-model reflects as a space right to the pore, see Fig. \ref{fig1} (dashed line) which size, one can change by means of the parameter $k$.

All sides of the outer space satisfy to the periodic boundary conditions. The left pore side reflects the inner molecules due to the boundary condition \cite{NPP1709} but also provides the periodic boundary conditions for a part of outer space.

There are $420$ molecules of water vapor inside the pore which form saturated water vapor at temperature $35^{\circ}$C and pressure $5.62\, kPa$ at the time $t=0$. The value of parameter $k = 3$ means that the outer space volume for calculations is $21$ times larger that the pore volume.  There are $1764$ molecules of water vapor in outer space  corresponding to 20\% saturated water vapor. The integration step is $0.016 ps$.

First, we made several runs of our program with 69 blocks of 32 threads. Each implementation took 2000 time steps. We found that performing  all the functions at every step in each run  is the same with a small deviation.  Time averages and deviations for each function,  as well as the overall time of one step are shown in table \ref{Tab1}.

\begin{table}[H]
\centering
\begin{tabular}{|c|c|c|c|c|}
\hline
 \multirow{2}{*}{F}&\multicolumn{2}{c|}{$t_{CPU}$ ($ms$)}&\multicolumn{2}{c|}{$t_{GPU}$ ($ms$)}\\
\cline{2-5}
  &$\bar{t}$&$\sigma_t$&$\bar{t}$&$\sigma_t$\\
\hline
F1&    112.575 &	2.372 &	4.479 & 0.110\\
\hline
F2+F3& 0.124	&0.005	&0.135	&0.009\\
\hline
F4& 0.038	&0.002	&0.054	&0.003\\
\hline
\end{tabular}
\caption{Calculation time averages $\bar{t}$ and deviations $\sigma_t$ for each function $F_i,i=1\div4$. $t_{CPU}$ -- \textbf{CPU} calculation time. $t_{GPU}$ -- total \textbf{GPU} calculation time.}    
\label{Tab1}
\end{table} 

For this reason, we will further consider that all program runs take an average implementation of time.

\begin{figure}[H]
\begin{center}
\includegraphics[width=0.8\linewidth]{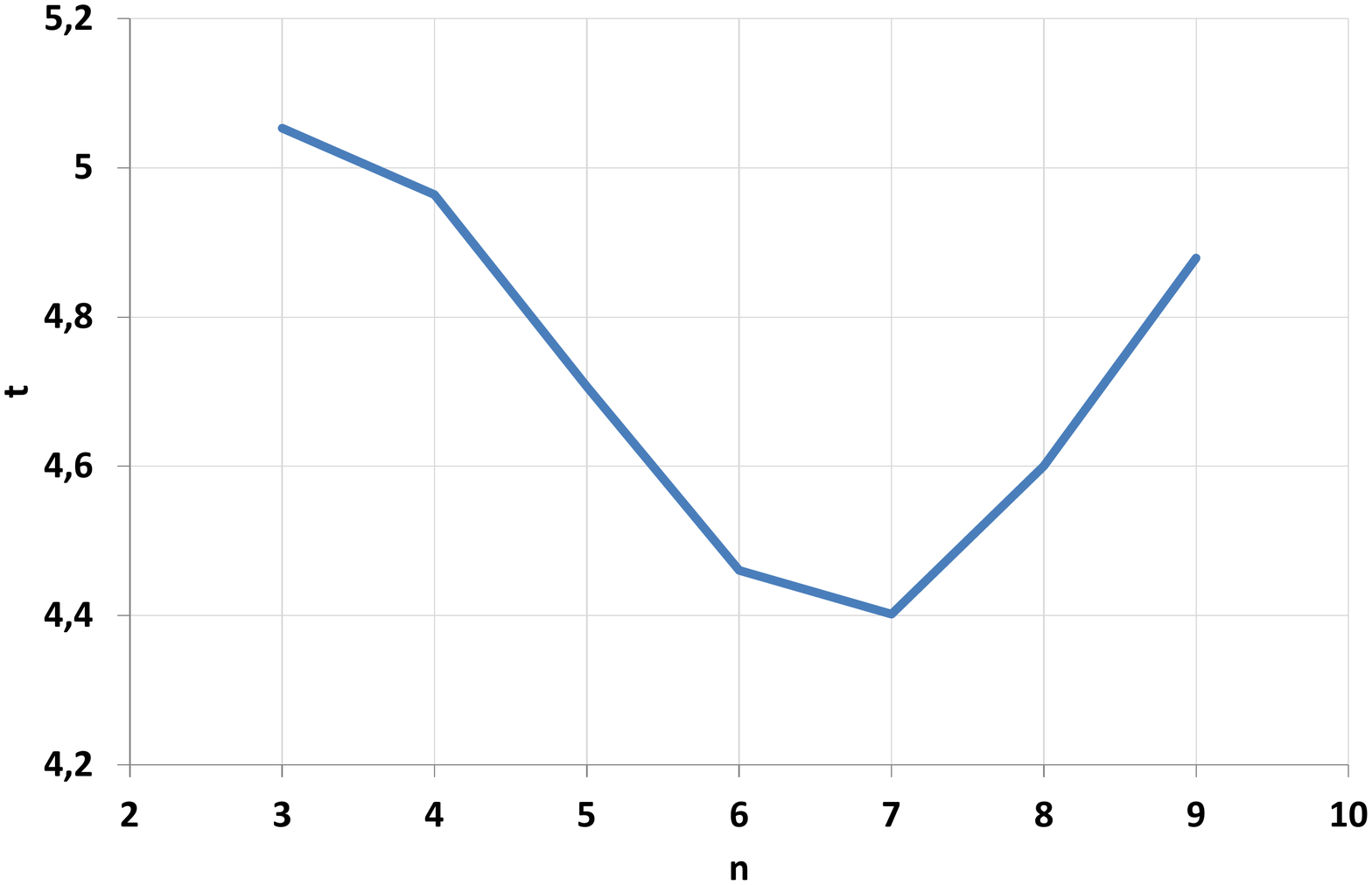}
\end{center}
\vspace{-1cm}
\caption{Dependence of the total GPU time calculation  on the number of threads per block. t  is the time in ms.  The number of threads in the block is $2^n$.}
\label{fig2}
\end{figure}

\begin{figure}[H]
\begin{center}
\includegraphics[width=0.8\linewidth]{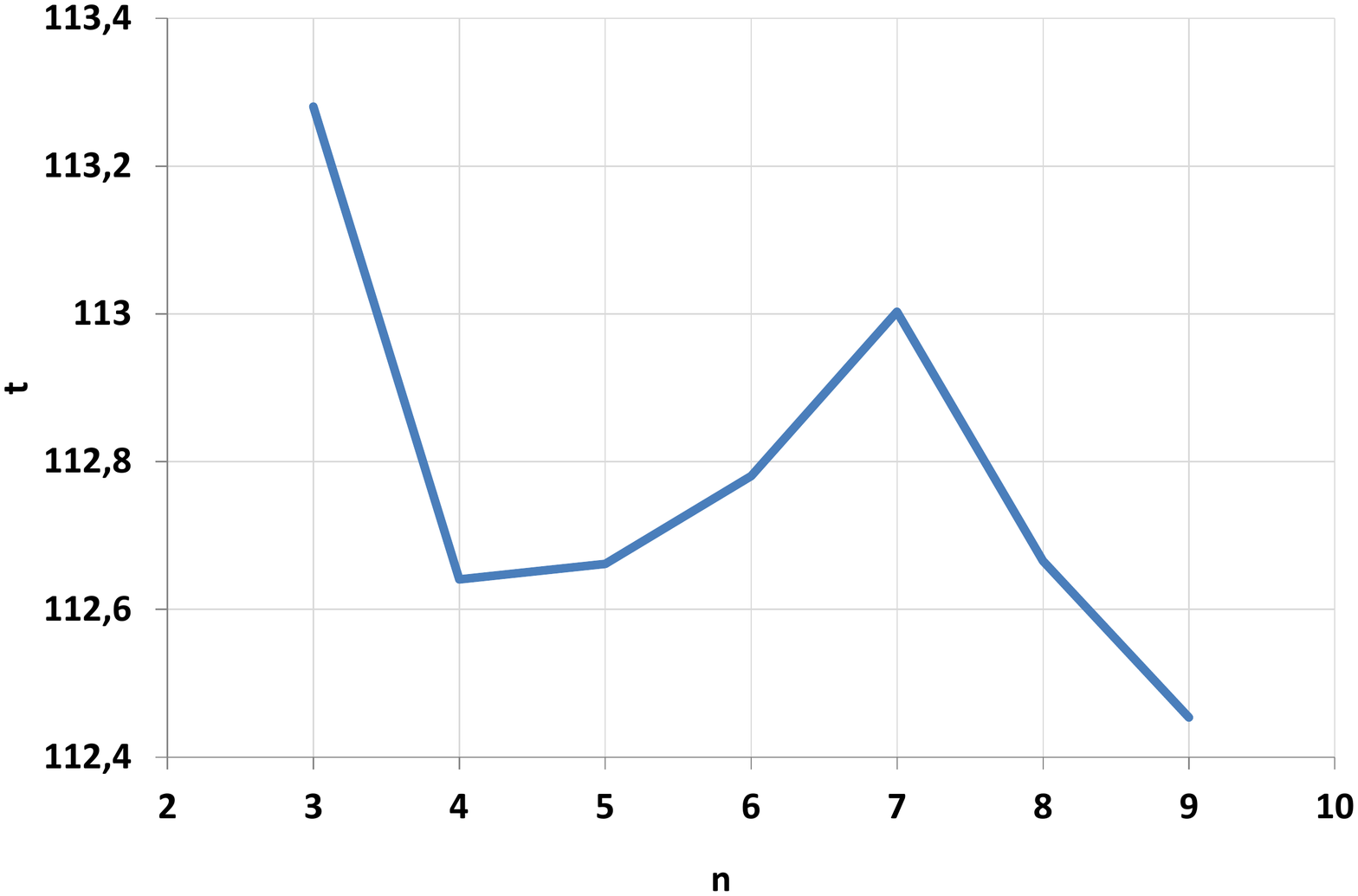}
\end{center}
\vspace{-1cm}
\caption{CPU calculation time dependence on the number of threads in a block. t  is the time in ms.  The number of threads in the block is $2^n$.}
\label{fig3}
\end{figure}

\begin{figure}[H]
\begin{center}
\includegraphics[width=0.8\linewidth]{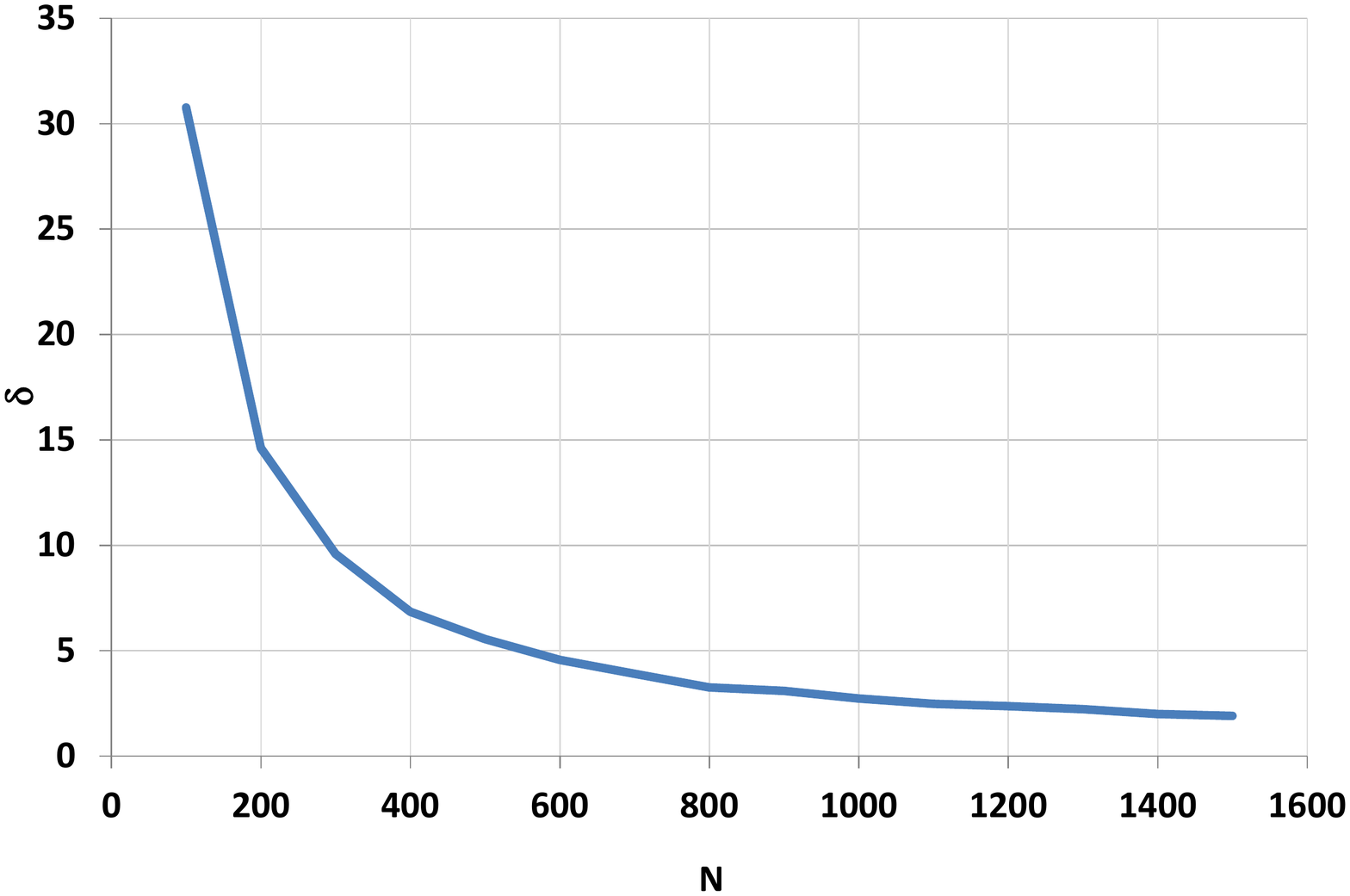}
\end{center}
\vspace{-1cm}
\caption{Comparison of the calculation time on the CPU and total GPU time depending on the number of particles in the pore expressed in percent $\delta = \frac{t_{GPU}}{t_{CPU}}\cdot 100\%$. $N$ is the number of particles in the pore.}
\label{fig4}
\end{figure}

\begin{figure}[H]
\begin{center}
\includegraphics[width=0.8\linewidth]{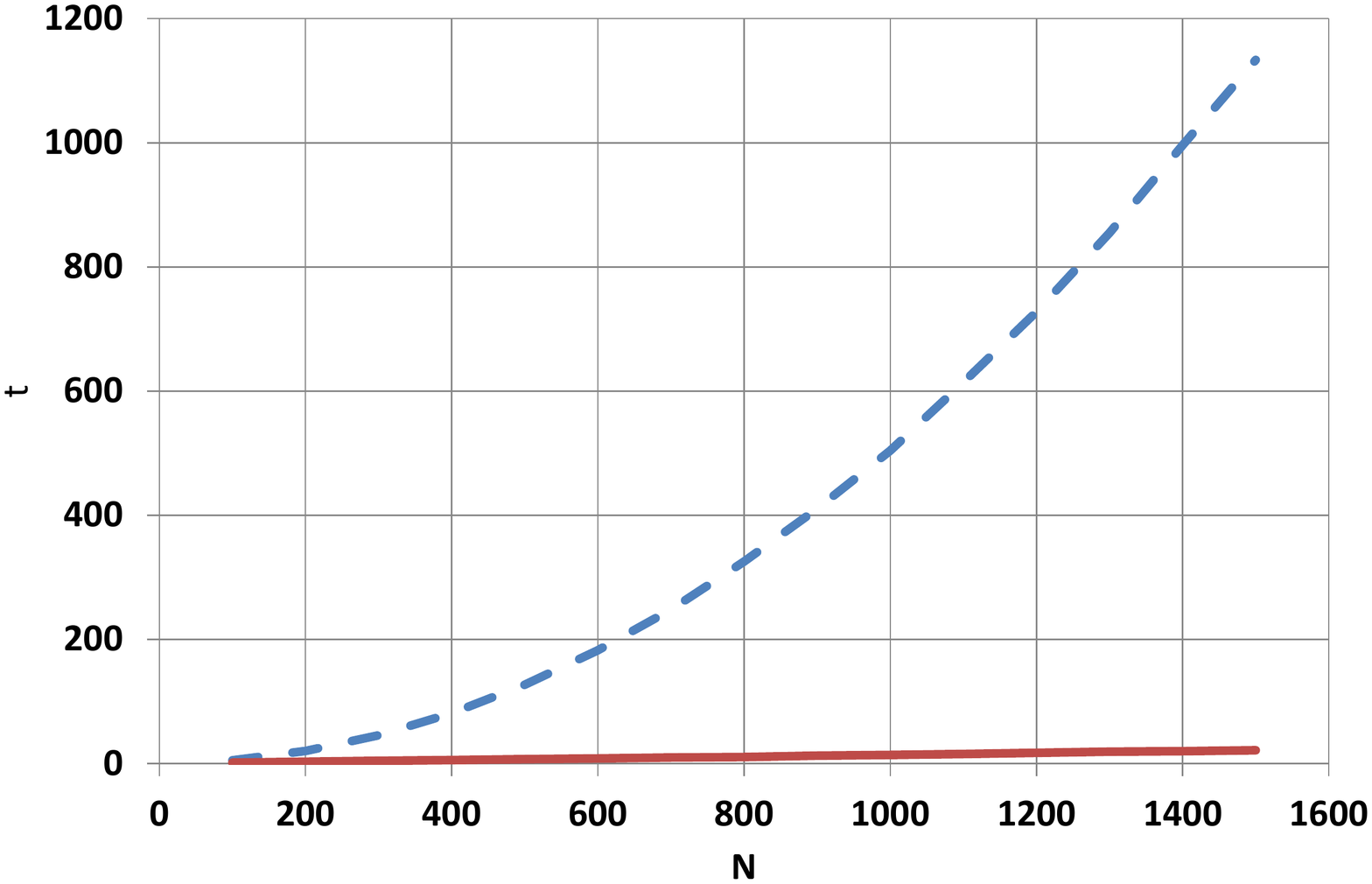}
\end{center}
\vspace{-1cm}
\caption{Comparison of computational time on the CPU and total GPU time depending on the number of particles in the pore. The dashed  line is the CPU calculation time and the solid line is the GPU calculation time.$N$ is the number of particles in the pore and $t$ is the time in ms.}
\label{fig5}
\end{figure}

\begin{figure}[H]
\begin{center}
\includegraphics[width=0.8\linewidth]{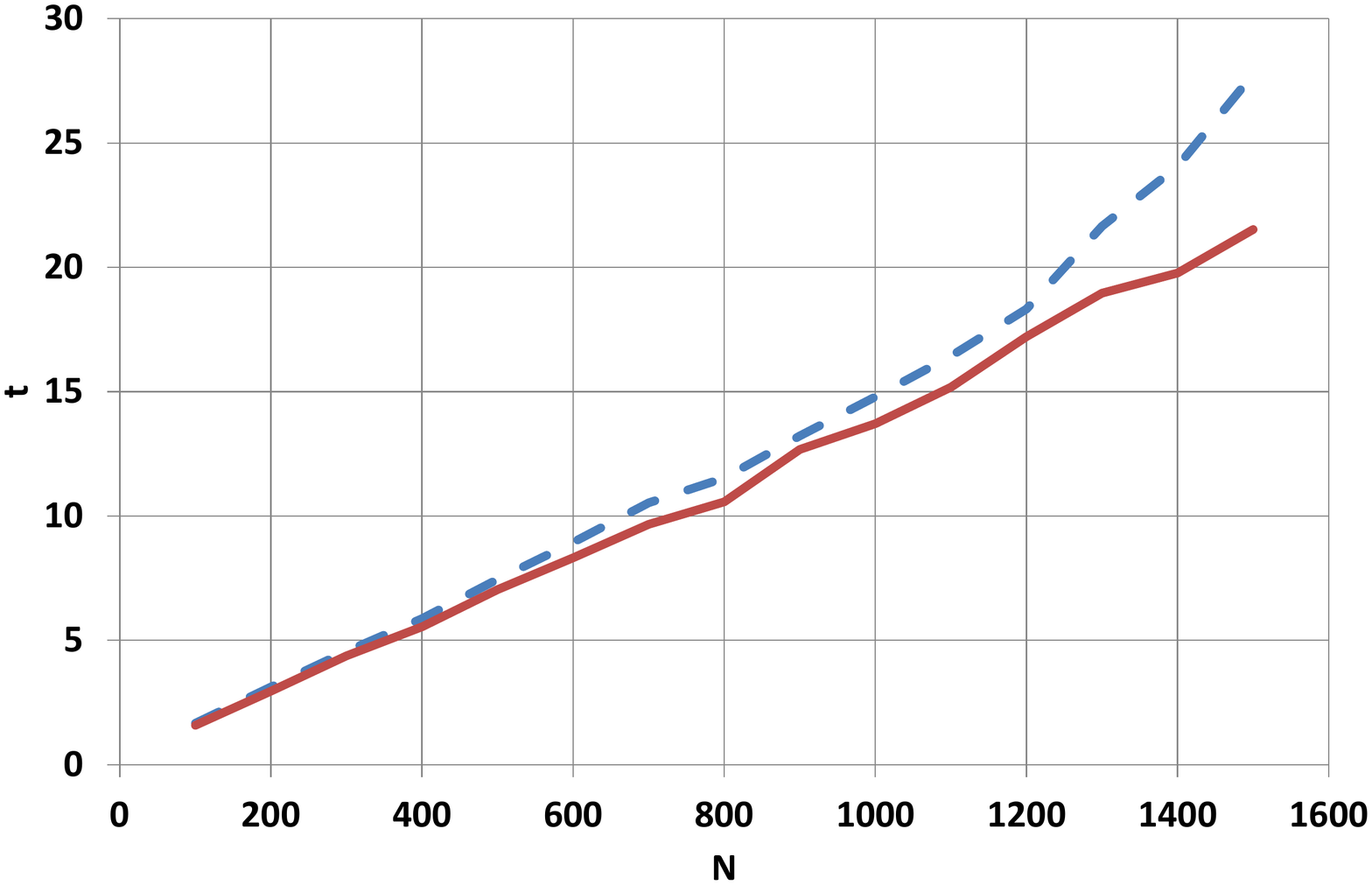}
\end{center}
\vspace{-1cm}
\caption{Total GPU calculation time for 128-threaded blocks (solid line) and 32-thread blocks (dashed line). $N$ is the number of particles in the pore and $t$ is the time in ms.}
\label{fig6}
\end{figure}

Furthermore, the total time of calculating the parallel portion of the code was examined, depending on the number of threads in the blocks (Fig. \ref{fig2}). We see that the minimum time has been reached for 128-threaded blocks (n = 7).  Such a dependency pattern did not have all 4 functions that are executed in parallel. The main creator of this result was a function to calculate the potential (F1). The calculation time on CPU varies within the calculated standard deviation (Fig. \ref{fig3}). When comparing CPU and GPU calculations time, we can see that GPU calculations were performed on average 24 times faster than CPU calculations.
Finally, we studied the calculation time for both platforms, depending on the number of particles in the pore while maintaining the ratio of the density in the pores and in the outer area of 5 : 1. It was used blocks with 128 threads. The results can be seen in table \ref{Tab2}.
\begin{table}[H]
\centering
\begin{tabular}{|c|c|c|c|}
\hline
$N$ & $t_{CPU}$ & $t_{GPU}$ & $\delta$ \\
\hline
100 &	5.125	&1.577	&30.775 \\
\hline
200	& 20.238	&2.957	&14.610 \\
\hline
300	& 45.588	&4.371	&9.589  \\
\hline
400	& 81.017	&5.541	&6.839  \\
\hline
500	& 126.854	&7.040	&5.550  \\
\hline
600	&182.441	&8.317	&4.559  \\
\hline
700	&248.107	&9.660	&3.894  \\
\hline
800	&325.447	&10.563	&3.246  \\
\hline
900	&410.330	&12.675	&3.089  \\
\hline
1000	&503.784	&13.706	&2.721  \\
\hline
1100	&614.173	&15.175	&2.471  \\
\hline
1200	&727.594	&17.212	&2.366  \\
\hline
1300	&855.108	&18.956	&2.217  \\
\hline
1400	&996.005	&19.767	&1.985  \\
\hline
1500	&1133.698	&21.522	&1.898  \\
\hline
\end{tabular}
\caption{The calculation time in dependence on the number of particles $N$ in the pore. Calculation time on CPU $t_{CPU}$ and total calculation 
time on GPU $t_{GPU}$ are given in ms. $\delta = \frac{t_{GPU}}{t_{CPU}}\cdot 100\%$.}  
\label{Tab2}  
\end{table}
On Fig.\ref{fig4}, we can see as it grows advantage of parallel computing  when the number of particles increases. Time needed for calculation on the CPU a total time  on the GPU is compared on figure \ref{fig5}. The development of the GPU calculation time for blocks with different  threads is shown on figure \ref{fig6}.

\section{3D molecular dynamic simulation}

In three-dimensional case\cite{NPP1708} we made simulation for a pore in the shape of a prism of dimensions  $l_x=500$ nm,\ $l_y=50$ nm,\ $l_z=50$ nm. Five walls are isolated and there is no exchange of particles with outer space. The sixth wall is open. The external environment is illustrated by a prism which is 9 times bigger than the pore. The big prism satisfies periodic boundary conditions. This means that the particles which pass through one wall return to the system through the opposite wall.  
Integration time step is $\Delta t=0.016$ ps and evolution time 65.3 ns.
For our purposes, we will again perform only 2000 time steps. 

Consequently, we have considered the following input data for the drying process: There are 1000 molecules of water vapor inside the pore which form saturated water vapor at temperature  $25$ $^oC$ and pressure $3.17$ $kPa$. There are 1800 molecules of water vapor in the outer area space  corresponding to 20\% saturated water vapor.

The simulation of this problem is solved using the \textbf{CUDA C} code  according to the computational scheme on Fig. \ref{VD_hyb}. Each of the 4 functions F1 - F4 is expanded to calculate the 3rd coordinate.

\begin{figure}[H]
\begin{center}
\includegraphics[width=0.8\linewidth]{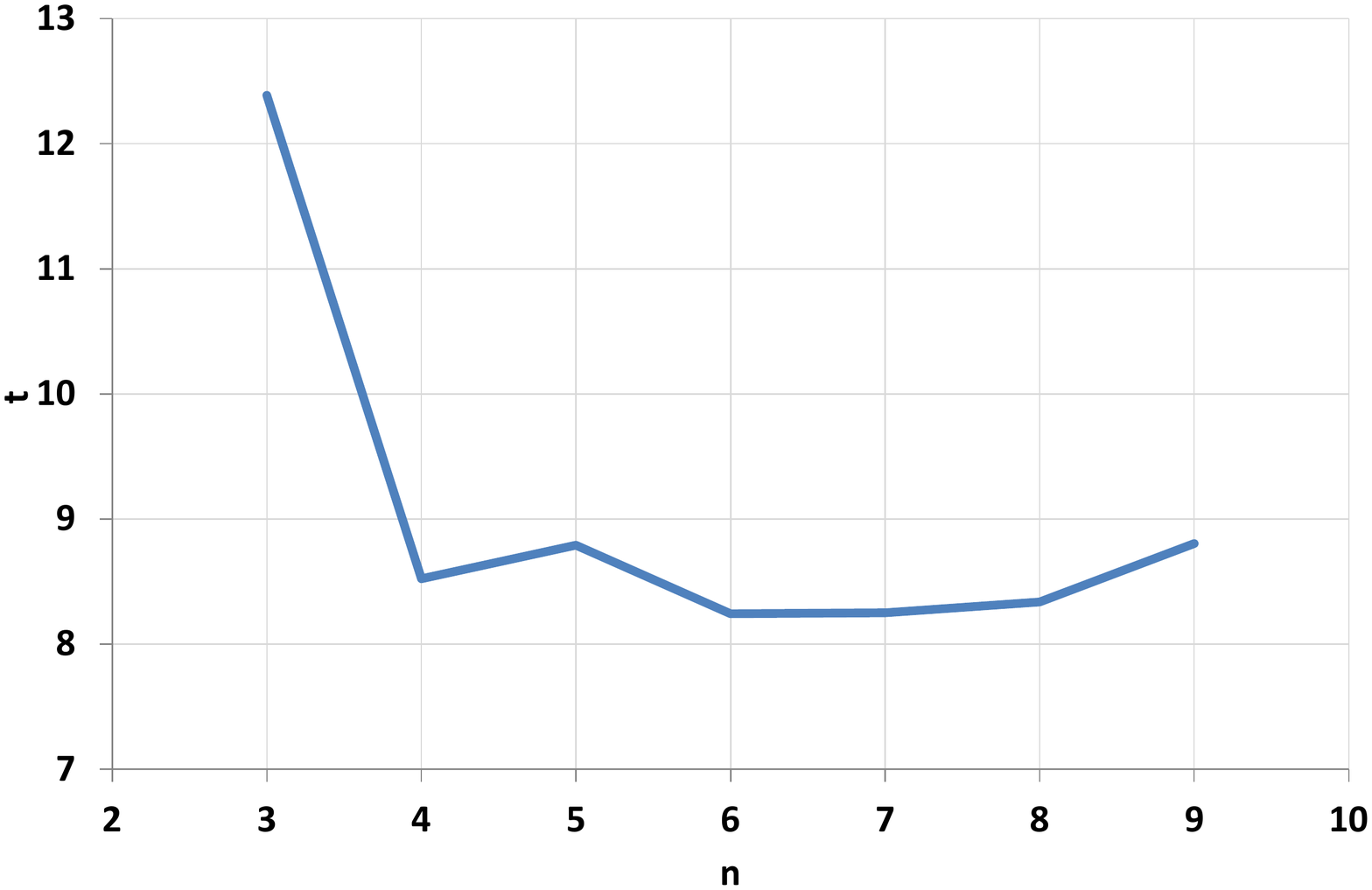}
\end{center}
\vspace{-1cm}
\caption{Dependence of the total GPU  time calculation on the number of threads per block for 3D simulation. $t$ is the time in ms. The number of threads in the block is $2^n$.}
\label{fig8}
\end{figure}

\begin{figure}[H]
\begin{center}
\includegraphics[width=0.8\linewidth]{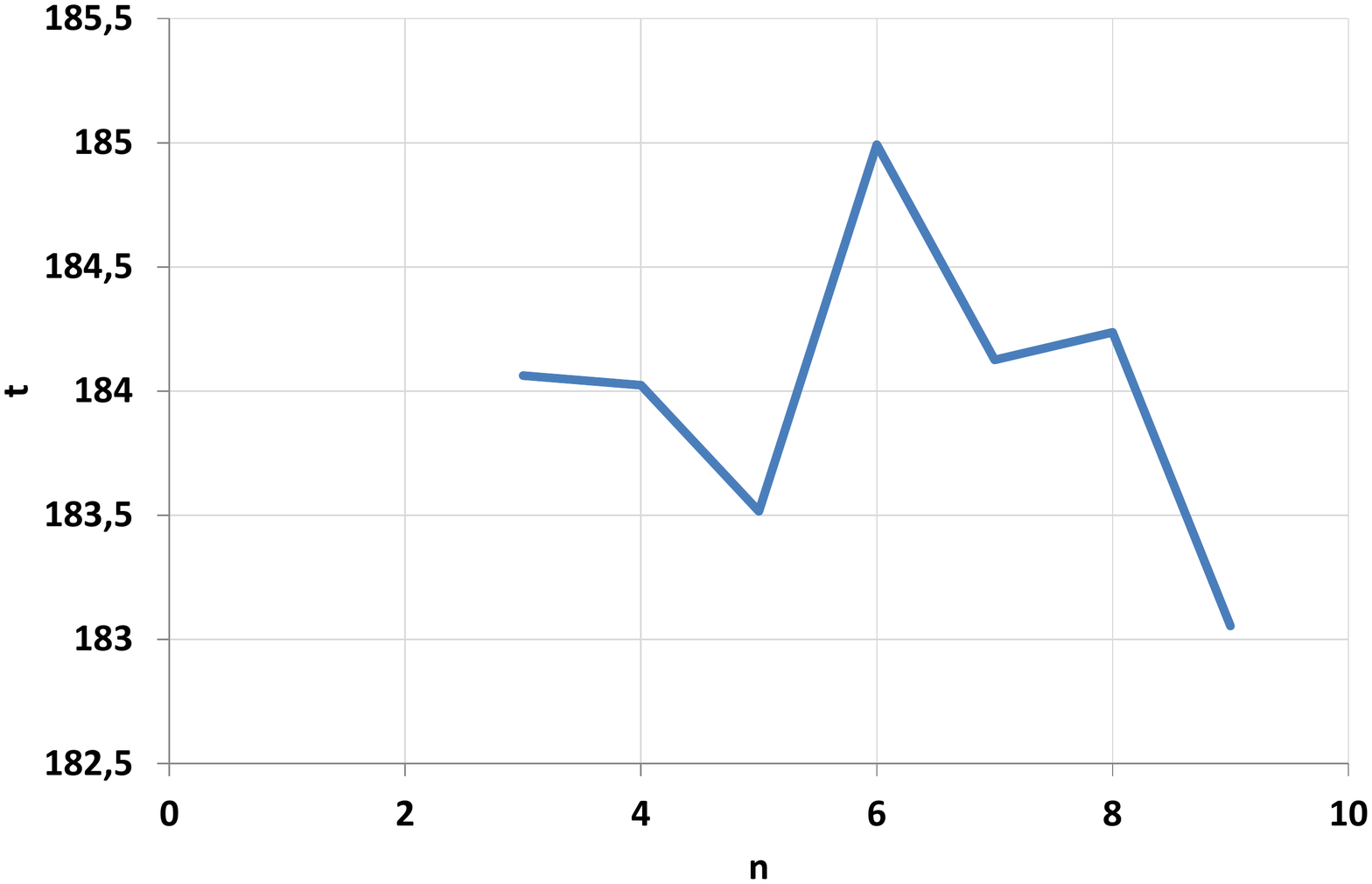}
\end{center}
\vspace{-1cm}
\caption{Development of computational time on the CPU depending on the number of threads in the block for 3D simulation. $t$ is the time in ms. The number of threads in the block is $2^n$.}
\label{fig9}
\end{figure}

We first look at the dependence of the total computational time for parallel computing on the number of threads in the block, this dependence can be seen on Fig. \ref{fig8}. The minimum time has been reached for blocks with 64, 128 and 256 threads (n = 6, 7, 8). Again the main creator of this result was a function to calculate the potential (F1). GPU calculations were performed on average 21 times faster than CPU calculations. Development of computational time on the CPU is shown on Fig. \ref{fig9}. 

Furthermore, the calculation time of both platforms was investigated depending on the number of particles in the pore. For 3D simulation, the same ratio of particles inside the pores and in outside was maintained like for 2D simulation. 128-threaded blocks were used for the calculations. The results are shown in Table \ref{Tab3}.

\begin{table}[H]
\centering
\begin{tabular}{|c|c|c|c|}
\hline
$N$ & $t_{CPU}$ & $t_{GPU}$ & $\delta$ \\
\hline
100 &	1.993	&1.042	&52.279 \\
\hline
200	&   7.417	&1.776	&23.944 \\
\hline
300	& 16.605	&2.504	&15.077  \\
\hline
400	& 29.462	&3.374	&11.453  \\
\hline
500	& 46.049	&4.115	&8.936  \\
\hline
600	&68.389		&4.829	&7.061  \\
\hline
700	&90.600		&5.571	&6.149  \\
\hline
800	&118.330	&6.162	&5.207  \\
\hline
900	&150.655	&7.276	&4.830  \\
\hline
1000	&184.301	&8.175	&4.436  \\
\hline
1100	&223.147	&9.035	&4.049  \\
\hline
1200	&263.869	&9.924	&3.761  \\
\hline
1300	&313.163	&10.501	&3.353  \\
\hline
1400	&360.352	&11.587	&3.215  \\
\hline
1500	&412.590	&12.434	&3.014  \\
\hline
\end{tabular}
\caption{The calculation time in dependence on the number of particles $N$ in the pore. Calculation time on CPU $t_{CPU}$ and total calculation 
time on GPU $t_{GPU}$ are given in ms. $\delta = \frac{t_{GPU}}{t_{CPU}}\cdot 100\%$.}    
\label{Tab3}
\end{table}

\begin{figure}[H]
\begin{center}
\includegraphics[width=0.8\linewidth]{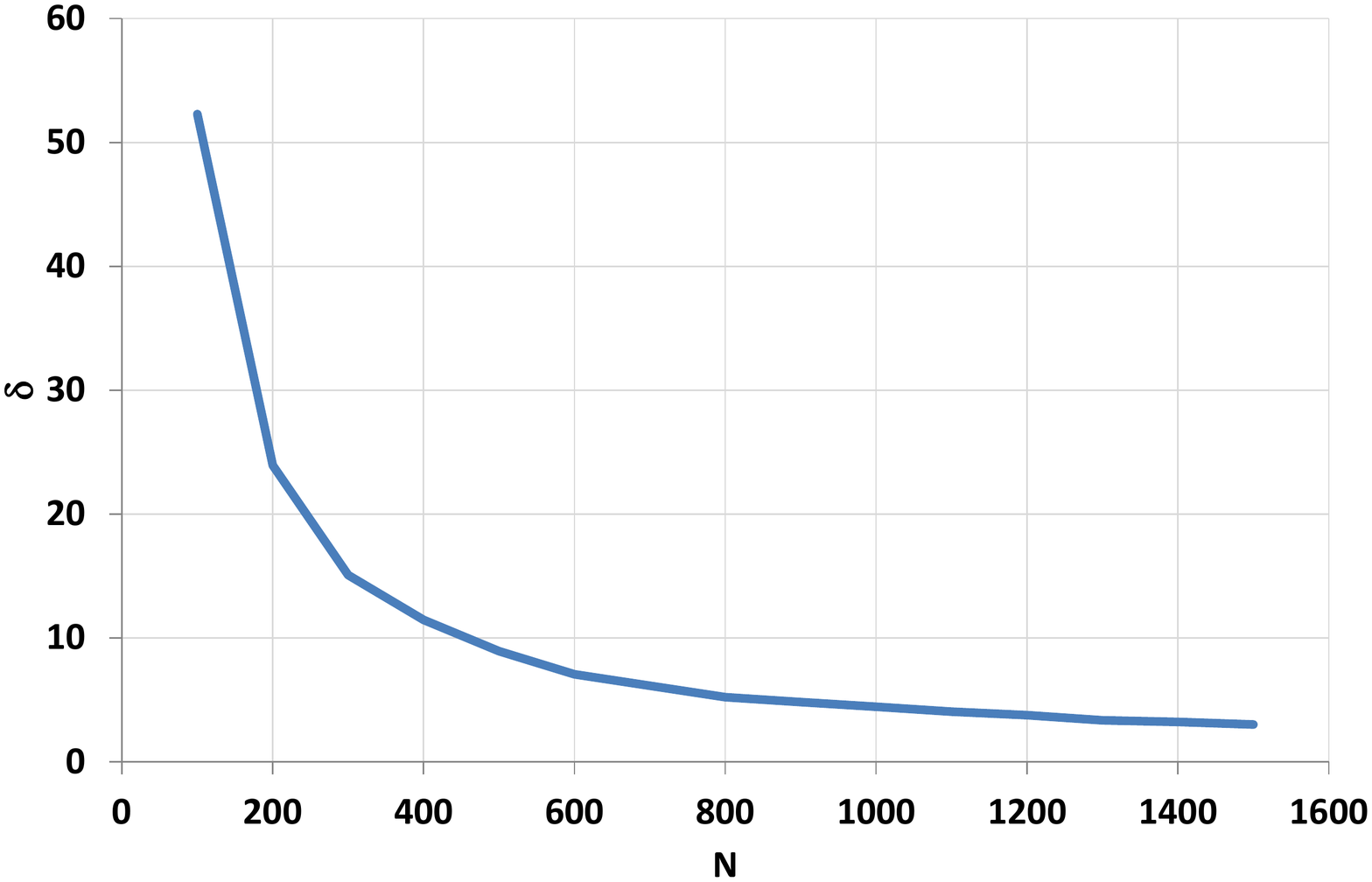}
\end{center}
\vspace{-1cm}
\caption{Comparison of the calculation time on the CPU and total GPU time depending on the number of particles in the pore expressed in percent $\delta = \frac{t_{GPU}}{t_{CPU}}\cdot 100\%$ for 3D simulation. $N$ is the number of particles in the pore.}
\label{fig10}
\end{figure}

\begin{figure}[H]
\begin{center}
\includegraphics[width=0.8\linewidth]{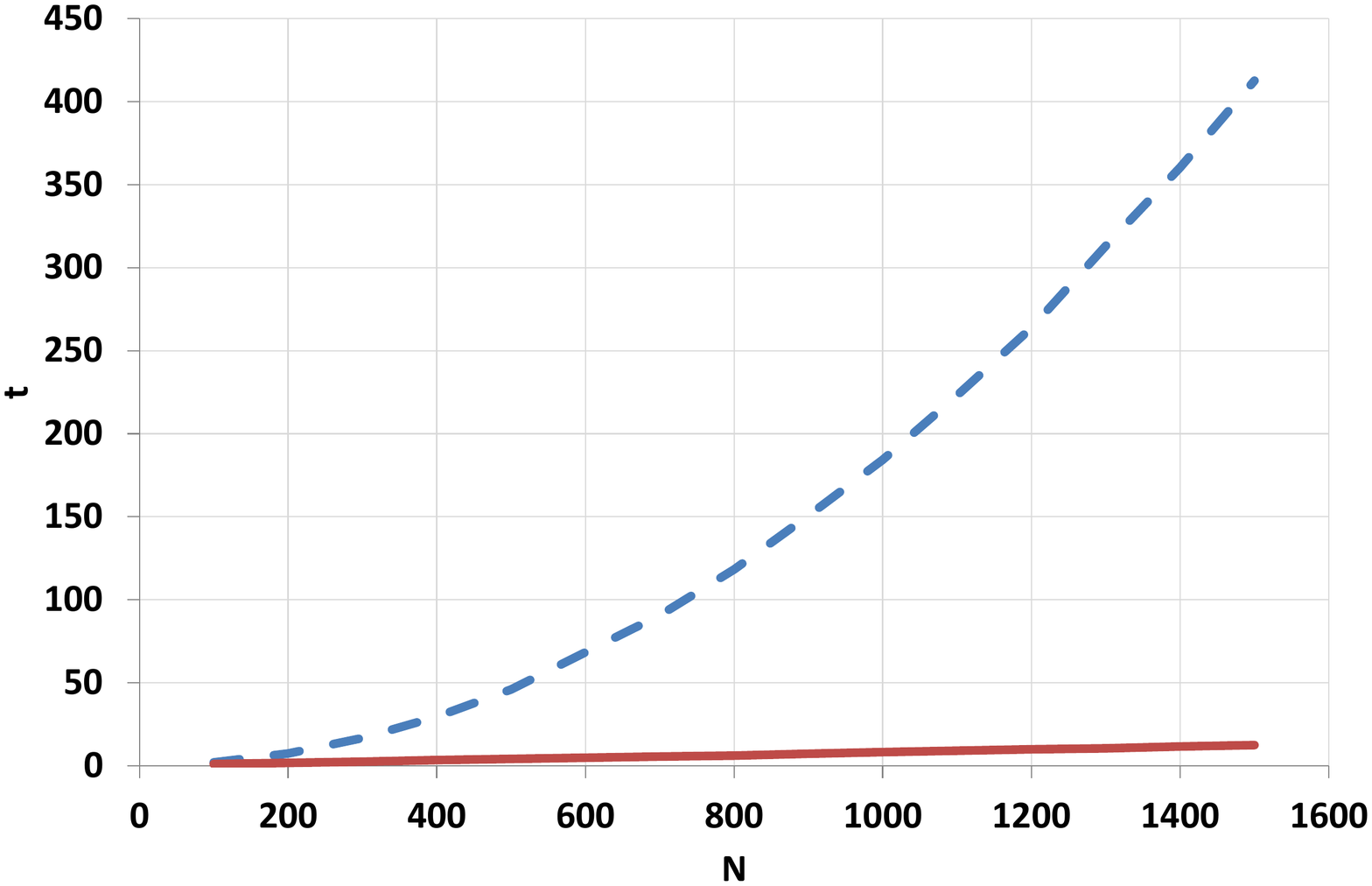}
\end{center}
\vspace{-1cm}
\caption{Comparison of computational time on the CPU and total GPU time depending on the number of particles in the pore. The dashed  line is the CPU calculation time and the solid line is the GPU calculation time. $N$ is the number of particles in the pore and $t$ is the time in ms.}
\label{fig11}
\end{figure}

\begin{figure}[H]
\begin{center}
\includegraphics[width=0.8\linewidth]{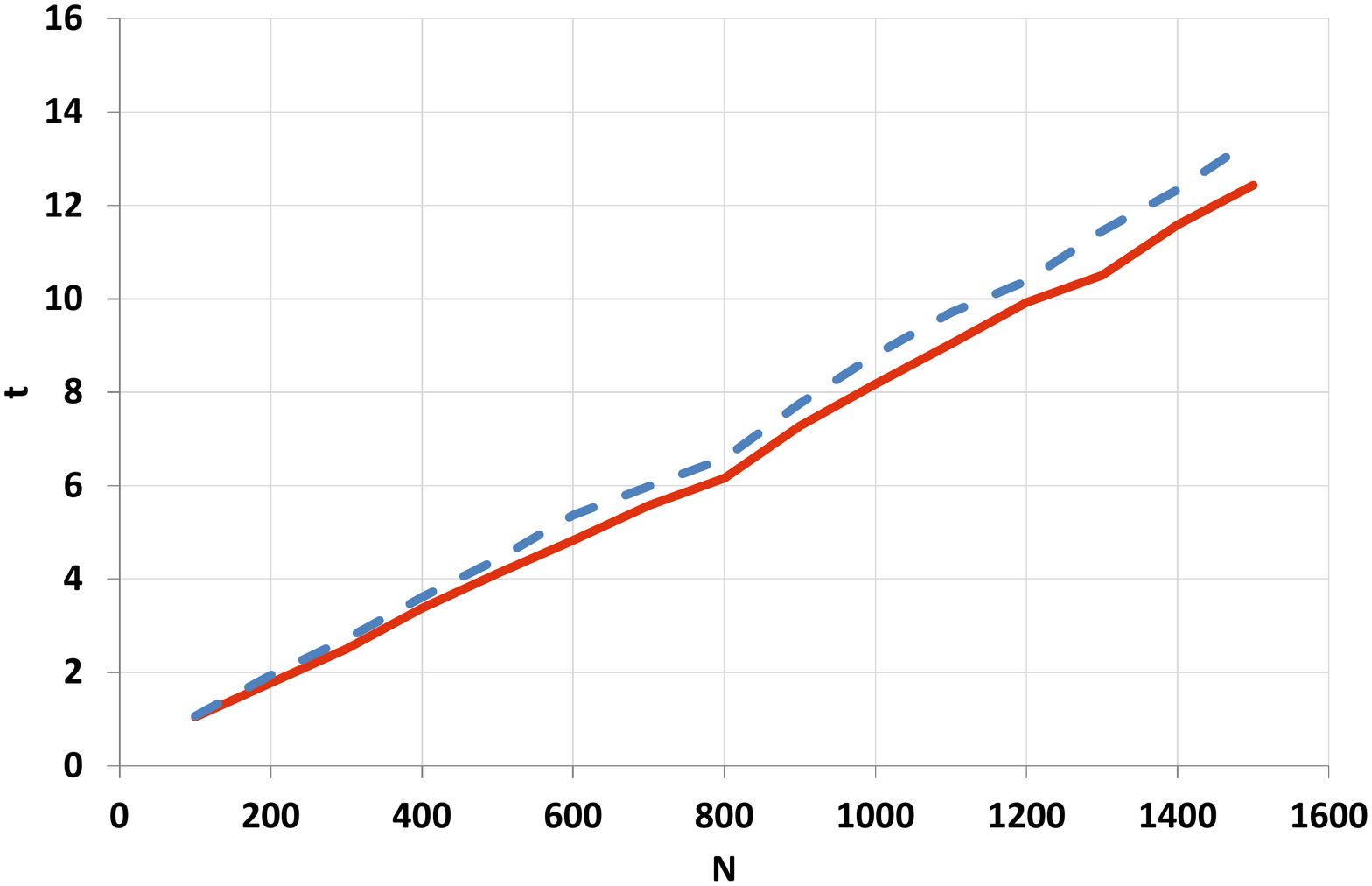}
\end{center}
\vspace{-1cm}
\caption{Total GPU calculation time for 128-threaded blocks (solid line) and 32-thread blocks (dashed line) for 3D simulation. $N$ is the number of particles in the pore and $t$ is the time in ms.}
\label{fig12}
\end{figure}

The advantage of parallel calculations for the increasing number of particles is shown on Fig. \ref{fig10}.  Comparison of CPU time and total GPU time is depicted on Fig. \ref{fig11}. The development of the GPU calculation time for blocks with different threads is shown on Fig. \ref{fig12}.

For both simulations, the average time required to calculate one step was also calculated. In order to compare the two simulations, we have converted this time to one particle. The results are shown in Table  \ref{Tab4}. The time required to calculate one step for one particle for 2D and 3D simulation on \textbf{GPU} in both cases (pure \textbf{GPU} time and total \textbf{GPU} time) especially pure \textbf{GPU} time is 2 : 3. \textbf{CPU} time does not keep this ratio.

\begin{table}[H]
\centering
\begin{tabular}{|c|c|c|c|c|c|c|}
\hline
 \multirow{2}{*}{time}&\multicolumn{3}{c|}{\textbf{2D}}&\multicolumn{3}{c|}{\textbf{3D}}\\
\cline{2-7}
  &CPU &pure GPU &total GPU &CPU &pure GPU &total GPU\\
\hline
One Step& 112.732	&4.532	&4.668	&184.349	&8.649	&8.786\\
\hline
Per Particle& 0.05162	&0.00207	&0.00214	&0.06584 &0.00309	&0.00314\\
\hline
\end{tabular}
\caption{Comparison of computation times per step and for one particle for both simulations.}    
\label{Tab4}
\end{table} 

\section{Conclusions}    
As our investigations showed for both cases of 2D and 3D simulation, when paralleling the computations, there are some optimal value of number of threads in blocks such that the computation time becomes minimal in comparison with other values of this number of threads. In addition, it should be noted that, when parallelizing, the cost ratio of the computation time per particle for 2D and 3D modeling is equal $2/3$ with high precision. 

\end{document}